\documentclass[5p,times,procedia]{elsarticle}
\usepackage{ecrc}
\volume{00}

\firstpage{1}

\journalname{Procedia CIRP}

\runauth{C. Ellwein et al.}

\jid{trpro}

\usepackage{multirow}
\usepackage[table]{xcolor}
\usepackage{tikz}
\definecolor{green}{RGB}{102,205,0} 
\definecolor{orange}{RGB}{255,165,0}     
\definecolor{red}{RGB}{255,99,71}
\newcommand{\colorcircle}[1]{%
  \tikz\draw[fill=#1, draw=#1] (0,0) circle (0.12cm);
}

\usepackage{amssymb}
\usepackage[figuresright]{rotating}

\usepackage[bookmarks=false]{hyperref}
    \hypersetup{colorlinks,
      linkcolor=blue,
      citecolor=blue,
      urlcolor=blue}

\usepackage{comment}
\usepackage{todonotes} %% activate todos

\begin{document}
\begin{frontmatter}

\dochead{CIRP Conference on Manufacturing Systems (CIRP CMS)}%

\title{Software-heavy Asset Administration Shells: Classification and Use Cases}

\author[a]{Carsten Ellwein\corref{cor1}} 
\author[a]{David Dietrich}
\author[a]{Jessica Roth} 
\author[b]{Rozana Cvitkovic}
\author[a]{Andreas Wortmann}

\address[a]{Institute for Control Engineering of Machine Tools (ISW), University of Stuttgart, Seidenstr. 36, D-70174 Stuttgart, Germany}
\address[b]{Blue Yonder GmbH, Industriestr. 6, D-70565 Stuttgart, Germany}

\cortext[cor1]{* Corresponding author. Tel.: +49-711-685-82424. {\it E-mail address:} carsten.ellwein@isw.uni-stuttgart.de}

%---------------------------------------------------------------------------------------------------
% 00_Abstract
%---------------------------------------------------------------------------------------------------

\begin{abstract}
% 1.) Introduction. In one sentence, what’s the topic? Phrase it in a way that your reader will understand. If you’re writing a PhD thesis, your readers are the examiners – assume they are familiar with the general field of research, so you need to tell them specifically what topic your thesis addresses. Same advice works for scientific papers – the readers are the peer reviewers, and eventually others in your field interested in your research, so again they know the background work, but want to know specifically what topic your paper covers.
The Asset Administration Shell~(AAS) is an emerging technology for the implementation of digital twins in the field of manufacturing.
% 2.) State the problem you tackle. What’s the key research question?  Again, in one sentence. (Note: For a more general essay, I’d adjust this slightly to state the central question that you want to address) Remember, your first sentence introduced the overall topic, so now you can build on that, and focus on one key question within that topic. If you can’t summarize your thesis/paper/essay in one key question, then you don’t yet understand what you’re trying to write about. Keep working at this step until you have a single, concise (and understandable) question.
Software is becoming increasingly important, not only in general but specifically in relation to manufacturing, especially with regard to digital manufacturing and a shift towards the usage of artificial intelligence. %Anmerkung EN: "The 2026 Conference Theme will be Driving the Future of Manufacturing Systems with AI and Sustainability."
This increases the need not only to model software, but also to integrate services directly into the AAS.
% 3.) Summarize (in one sentence) why nobody else has adequately answered the research question yet. For a PhD thesis, you’ll have an entire chapter, covering what’s been done previously in the literature. Here you have to boil that down to one sentence. But remember, the trick is not to try and cover all the various ways in which people have tried and failed; the trick is to explain that there’s this one particular approach that nobody else tried yet (hint: it’s the thing that your research does). But here you’re phrasing it in such a way that it’s clear it’s a gap in the literature. So use a phrase such as “previous work has failed to address…”. (if you’re writing a more general essay, you still need to summarize the source material you’re drawing on, so you can pull the same trick – explain in a few words what the general message in the source material is, but expressed in terms of what’s missing)
The existing literature contains individual solutions to implement such software-heavy AAS. 
% 4.) Explain, in one sentence, how you tackled the research question. What’s your big new idea? (Again for a more general essay, you might want to adapt this slightly: what’s the new perspective you have adopted? or: What’s your overall view on the question you introduced in step 2?)
However, there is no systematic analysis of software architectures that integrate software services directly into the AAS.
% 5.) In one sentence, how did you go about doing the research that follows from your big idea. Did you run experiments? Build a piece of software? Carry out case studies? This is likely to be the longest sentence, especially if it’s a PhD thesis – after all you’re probably covering several years worth of research. But don’t overdo it – we’re still looking for a sentence that you could read aloud without having to stop for breath. Remember, the word ‘abstract’ means a summary of the main ideas with most of the detail left out. So feel free to omit detail! (For those of you who got this far and are still insisting on writing an essay rather than signing up for a PhD, this sentence is really an elaboration of sentence 4 – explore the consequences of your new perspective).
This paper aims to fill this research gap and differentiate architectures based on software quality criteria as well as typical manufacturing use cases.
% 6.) As a single sentence, what’s the key impact of your research? Here we’re not looking for the outcome of an experiment. We’re looking for a summary of the implications. What’s it all mean? Why should other people care? What can they do with your research. (Essay folks: all the same questions apply: what conclusions did you draw, and why would anyone care about them?)
This work may be considered as an interpretation guideline for software-heavy AAS, both in academia and for practitioners.
\end{abstract}

\begin{keyword}
Digital Manufacturing \sep Digital Twin \sep AAS
\end{keyword}

\end{frontmatter}

%---------------------------------------------------------------------------------------------------
% 01_Introduction
%---------------------------------------------------------------------------------------------------
\section{Introduction}
\label{sec:Introduction}

For digital manufacturing, the Asset Administration Shell (AAS)~\cite{wei2019review} is a potential standard for modeling and implementing digital twins (DTs)~\cite{zhang2025} of products~\cite{ajdinovic2024} processes~\cite{dietrich2024} and resources~\cite{chen2025} 
in the context of production~\cite{ellwein2023} and related disciplines, such as civil engineering~\cite{ellwein2021}.
The current specifications of the AAS define a metamodel and use case-specific submodels for data representation, as well as an application programming interface (API) with its security guidelines.
The expansion stages defined therein, commonly known as AAS types, suggest communication with the asset and associated services, which lift the AAS from a digital model of the related asset to an autonomous and service-integrated DT with dynamic context-aware behavior.
Thus, the software component gains importance in digital manufacturing with the need to represent the software services itself in the AAS~\cite{ellwein2023, frick2024}.

In addition to AASs that contain logic in software services related to, but not contained in the AAS \cite{Stolze.2024}, first proof of concepts are realized that implement the software itself in the AAS \cite{Ellwein.2025a}.
With regard to managing the DT's software or services of varying complexity or size in the AAS, there are currently no sufficient concepts in the state-of-the-art and specifications mentioned, as the following related works show.
This paper addresses this gap by contributing a classification of software-heaviness regarding the runtime environment and functional components.
Out of this concept possible architectures with their effects on software quality criteria and corresponding use-cases are derived.
Our contribution thus enables the further development of AASs with varying software heaviness in accordance with the standard to realize use cases in digital manufacturing, e.g., a preprocessing of incoming data from the asset or performing actions out of the AAS.
To achieve that, this paper is structured as follows: \autoref{sec:RelatedWork} reviews the related work of software services in DTs and AASs.
The classification of the software-heaviness of AAS is introduced in \autoref{sec:Classification}.
In \autoref{sec:SystemDesign} architectures, quality criteria and use cases are evolved.
The paper concludes in \autoref{sec:Conclusion} with a summary and perspective of the findings.

%---------------------------------------------------------------------------------------------------
% 02_RelatedWork
%---------------------------------------------------------------------------------------------------
\section{Related Work}
\label{sec:RelatedWork}
%--------------------
% DT
%--------------------

\begin{figure} [t]
\centering
    \includegraphics[width=1\columnwidth]{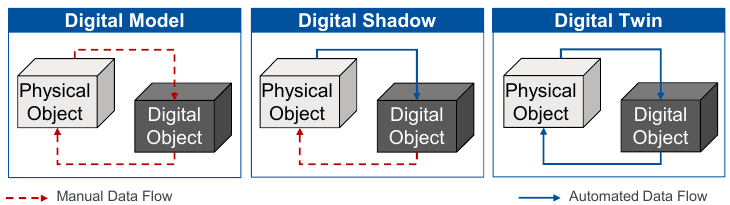}
    \caption{Distinguishing between digital model, shadow, and twin based on the nature of their data interactions with the real system, based on \cite{kritzinger2018} }
    \label{fig:kritzinger}
\end{figure}

DTs are a digital representation of an actual cyber-physical (production) system \cite{eramo2021}.
A distinction is made between three levels of automated data flows between physical objects and digital representations \cite{kritzinger2018} (cf. \autoref{fig:kritzinger}): 
(i)~A \textit{digital model} represents a physical object; any change in the physical or digital counterpart requires manual interaction to ensure consistency. 
(ii)~A \textit{digital shadow} (DS) automatically reflects changes in the physical object through an interface. 
(iii)~In a \textit{DT}, the physical and digital objects (e.g., model, DS) are fully integrated and the information is automatically fully synchronized. 
In order to implement digital manufacturing use cases, a software component of the DT must have access to a sufficient information model of the plant. 
This is essential for seamlessly retrieving information about the physical entity and enabling interoperable interaction with other DTs \cite{dalibor2022cross}.

\begin{figure}[t]
    \centering \includegraphics[width=.85\columnwidth]{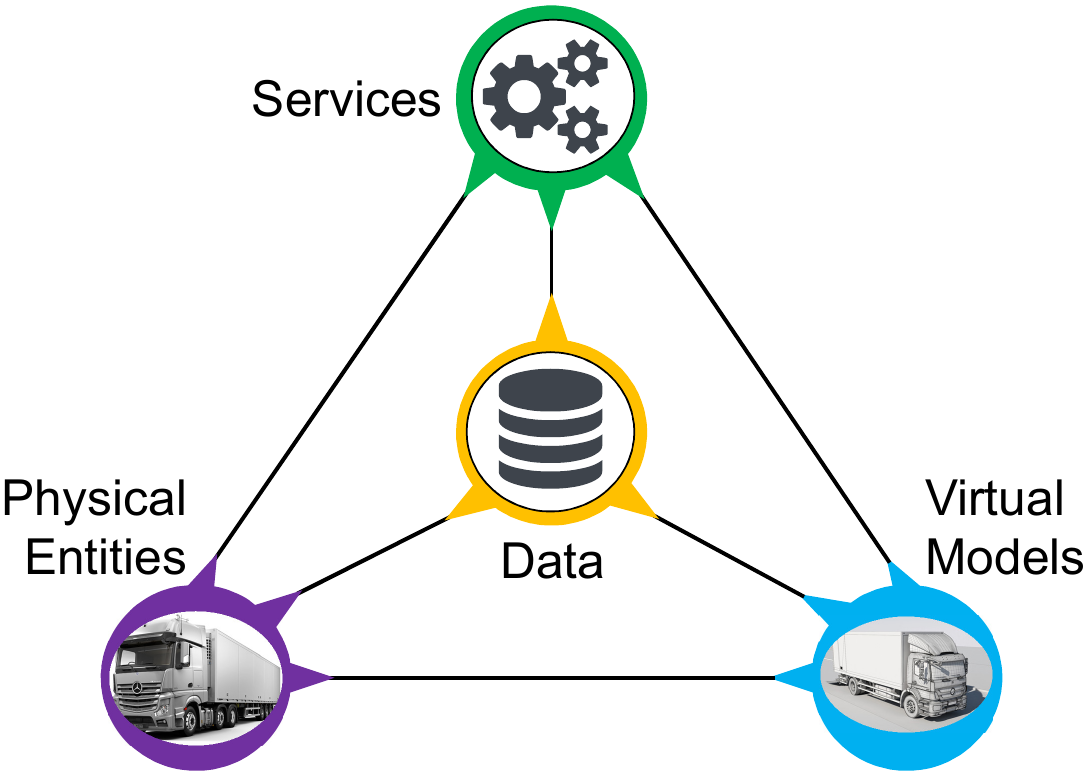}
    \caption{Digital twins based on their components and data flows, based on~\cite{tao2019}}
    \label{fig:tao5D}
\end{figure}

The DT model is defined by data flows \cite{kritzinger2018}, which emphasizes the connection between physical objects and their digital representations, supplemented by an internal perspective of the DT components. 
According to \cite{tao2019} (cf. \autoref{fig:tao5D}), a DT consists of: (i)~\textit{data}, (ii)~\textit{virtual models}, (iii)~\textit{services}, (iv)~\textit{physical entities}, and (v)~the \textit{connection} between the components of the DT mentioned above.

%--------------------
% AAS
%--------------------

One technology for implementing the digital object of a DT is the AAS \cite{zhang2025}. 
The specification of the AAS provides a distinction into three levels of enhancement, known as AAS types~\cite{AASReadingGuide}.
The AAS type~1 is consistently described as a static or passive representation of information about assets~\cite{13, 326}.
In contrast, stakeholders  interpret the types~2 and 3 differently~\cite{ellwein2026}.
Type~2 is described as AAS, providing a common interface or standardized API to access data~\cite{13, 326} or capable of responding to external events~\cite{13, 38}, but also as AAS with decision-making functionalities~\cite{88} or AAS including capabilities and skills~\cite{87}.
Type~3 is described as AAS, interacting with other type~3 AAS~\cite{38}, as well as AAS communication according to the I4.0 language~\cite{326}, as AAS containing service-oriented communication~\cite{87}, or AAS containing decision making skills~\cite{13}.
In summary, types~2 and 3 exchange data with the physical object. 
Hence, the importance of software as part of the AAS increases.
To provide greater clarity about the intended message, a portfolio analysis is proposed as an alternative to type designation, which distinguishes between the level of communication and the level of integration~\cite{ellwein2026}. 

Although many applications use the AAS for data exchange, few works consider software as part of the AAS.
The work can be divided into two different categories: (i) the literature that extends the architecture of the Eclipse BaSyx\footnote{https://basyx.org/} framework and (ii) the literature that uses container technology to extend the functionality of the AAS.
However, it should be noted that the BaSyx framework is also used in the second category for implementation, although the proposed concepts are widely decoupled from the BaSyx architecture.
Works considering software as part of the AAS with build on other architectures like FA3ST\footnote{https://www.iosb.fraunhofer.de/en/projects-and-products/faaast-ecosystem-digital-twins-aas.html} have not been found.
Within the first category (i), the division of functionalities between the AAS server and separate applications to realize type~3 AAS is discussed~\cite{Grunau.2022b}.
Furthermore, peer interaction is performed for a production planning scenario using Node-RED and a Business Process Modeling Notation (BPMN)~\cite{chinosi2012bpmn} engine with extension to the AAS repository~\cite{Stolze.2024}.
Within the second category (ii), scalable peer interaction is presented using a standalone AAS that includes data, logic, and communication capabilities in a container~\cite{Dietrich.inpress}.
In turn, orchestration through a separate AAS submodel and execution from within the AAS are also evaluated to realize service capabilities~\cite{Ellwein.2025a}. 

As can thus be determined, DTs represent complex systems that require bidirectional data exchange between the physical and digital layers, as well as the integration of services and logic.
The AAS provides a standardized framework for this purpose, but is implemented in various ways across the literature.
Although many approaches use the AAS primarily as a data container or interface, only a few works integrate software components directly into the AAS structure.
A unified classification that outlines which forms of software integration are possible and appropriate for different use cases is still missing.

%---------------------------------------------------------------------------------------------------
% 03_Classification
%---------------------------------------------------------------------------------------------------
\section{Classification}
\label{sec:Classification}
To determine the software heaviness of AAS, the runtime environment and the functional components must be differentiated into categories.
With this goal in mind, two multi-level AAS classifications are introduced within the following sections. 
These two classifications are independent of each other and should not be confused with the AAS (communication)~types (cf.  ~\cite{AASReadingGuide,ellwein2026}). 
The classifications introduced do not claim to be absolutely correct or complete, but are primarily intended to provide common ground for \autoref{sec:SystemDesign}. 
Although the classification of the runtime environment mainly reflects the usage phase of the software service, the functionality is an inherent part of the AAS in the whole lifecycle.

\subsection{Runtime Environment}
\label{subsec:Classification-Runtime}
In this context, the runtime environment of an AAS can be implemented in three different stages.
A passive variant provides an AAS with no additional runtime environment for service execution, e.g. stored as a file including.
The AAS only serves as a knowledge store, e.g. for sharing. Even if services were present in the AAS, it would not be possible to execute them. 
In terms of Kritzinger's definition, this AAS would correspond to a digital model, as neither automated dataflows to the asset nor corresponding services are present \cite{kritzinger2018}.

When an identical AAS is uploaded to a platform, like, e.g. BaSyx server, the server-based runtime environment is used. 
The server provides a generic runtime that can be used equally by all AAS on the server and offers generic features like the implementation of the AAS REST API and security according to its specification, eventing, operation delegation, the execution of scripts, or the other.
If machine communication is implemented, the runtime environment could update the AAS model or start operations, creating an AAS with communication and integration capabilities, which would be equivalent to a DS or a DT in Kritzinger's definition \cite{kritzinger2018}. 
In complex systems, this strategy could be used as it enables the use of middleware and additional services to be set up at scale, instead of setting up individual runtimes for AASs.

The third strategy describes the compilation of the AAS model and the runtime environment. 
The resulting standalone application can be consolidated from the server-based deployment and consists, for example, of the BaSyx SDK, the AAS model, and the runtime environment. 
The result is a ready-to-use executable, portable, self-contained application.
The runtime environment as a software component is more valuable and mature in comparison between the other two strategies, while the complexity also increases.
A standalone version may offer the same runtime functionalities as the server-based runtime. 
In extension, the standalone runtime offers the possibility to include high specific and encapsulated software functionalities.

It is theoretically possible to switch between these strategies by changing the runtime of the AAS since the AAS itself is independent of the runtime environment and is used only in different ways.
For example, by exporting the AAS model from the BaSyx server from server-based to file-based without runtime.
There are use cases in which the change is not advisable or can only be carried out at great expense or with loss of functionality at use.
Like that, to switch from a standalone AAS to a server-based or file-based AAS requires implemented serialization methods and may lose functionality that was provided by the standalone runtime.
The choice of runtime environment should be made based on the requirements of the project and quality characteristics such as performance and reliability, rather than on the type of AAS.

In summary, there are a descriptive model and a runtime environment, i.e. a system that handles requests and data processing. 
For the end user, it appears to function as an AAS, regardless of which architecture was used. 
This achieves an extension of the paradigm, as a distinction between model and environment has been created within the system, but it acts as a single unit externally and is therefore not visible to the user.

\subsection{Functional Components}
\label{subsec:Classification-Components}
\begin{figure}[t]
    \centering
    \includegraphics[width=0.95\linewidth]{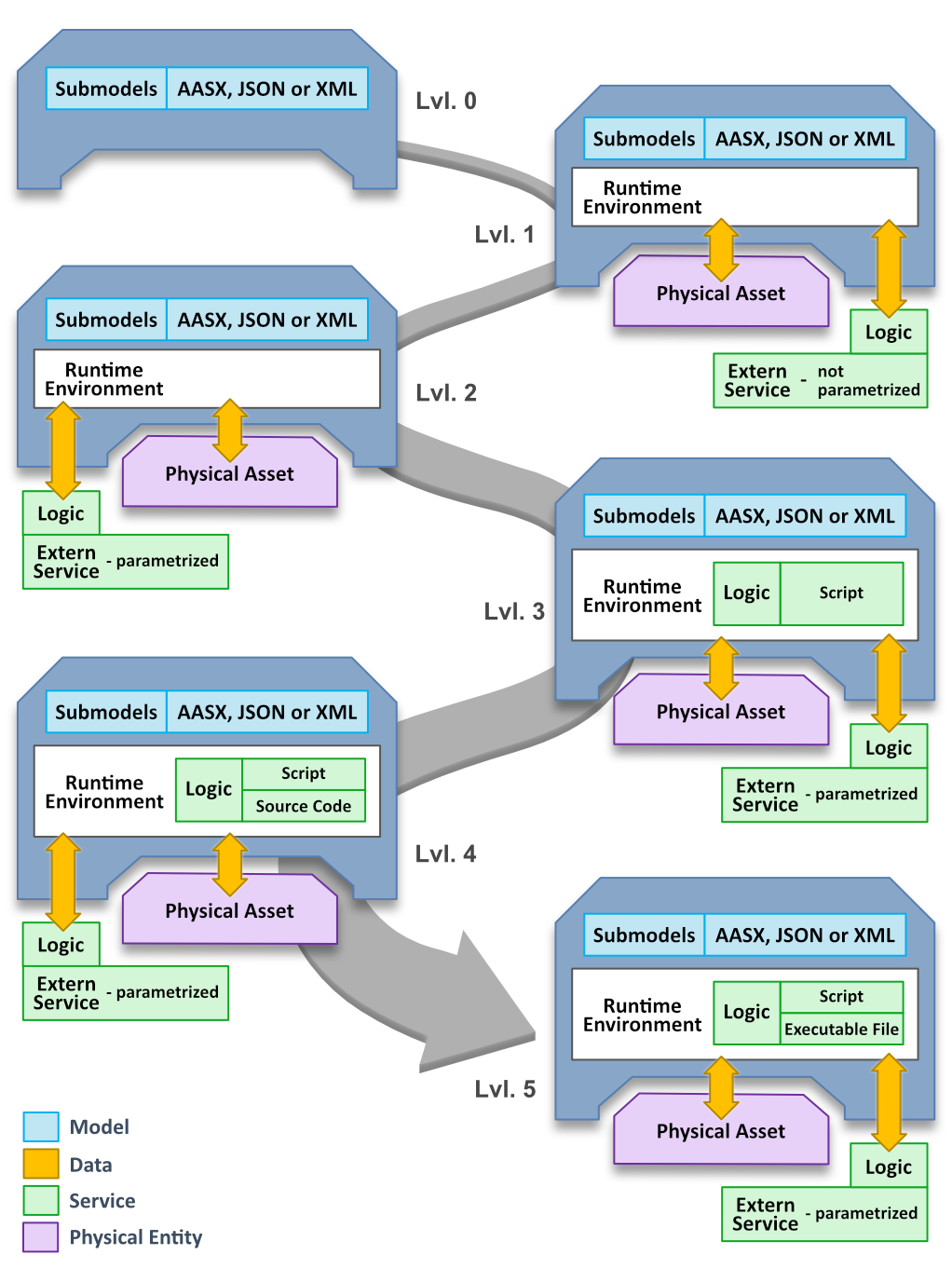}
    \caption{Six levels of software-heavy AAS}
    \label{fig:functionalComponents}
\end{figure}

After explaining the basic runtime strategies, the degree of software heaviness of AASs is examined. 
Six levels are used for classification, as shown in \autoref{fig:functionalComponents}. 
For structuring the components of an AAS, the 5D model \cite{tao2019} is used (cf. \autoref{fig:tao5D}).
The AAS model and data without functional services are referred to as level~0~(Lvl.~0) with respect to its software heaviness. 
Regarding the 5D model, the services are not present and thus also an automated connection to the AAS's physical entity is missing.
In the smallest extent of software functionality, called level~1~(Lvl.~1), the services enable a connection to the corresponding physical object and external logic.
For it's use a runtime environment is required.
Simple requests without specific parameterization are possible via the API, e.g. queries via the host name or simple OPC-UA parameters. 
Compared to Lvl.~0, the initial simple functionalities are provided within the system.
This enables external services to access the AAS data.
According to the 5D model, all five components can be present from this level on with increasing complexity. 
Level~2~(Lvl.~2) differs from the previous level in that requests with parameterization are possible. 
This allows data to be requested specifically via the service interface, and complex queries are possible.
Examples are requests via a REST interface or OPC-UA functions. 
The model must have knowledge of the interface to process requests accordingly.

Starting with level~3~(Lvl.~3), functionality is provided directly within the AAS. 
The logic of the service is no longer part of the periphery but is now located within the AAS itself. 
The service is described in scripts, containers, or regular expressions and is interpreted and executed by the system.
Level~4~(Lvl.~4) is an extension of Lvl.~3 in which the service is now described by source code rather than scripts, which enables increasing complexity of the service.
The system translates and executes the service within the AAS. 
This gives opportunities to optimize performance.
In level~5~(Lvl.~5), the AAS owns the complete service as an executable file.
This enables the system to directly execute the service from the AAS without interpreting or compiling it first.
In these levels, the software-heaviness of the AAS is increased stepwise with each level.
The logic of the service is included in the AAS, and the software components also become more complex, which leads to increasing requirements regarding its runtime environment.

%---------------------------------------------------------------------------------------------------
% 04_Design
%---------------------------------------------------------------------------------------------------
\section{System Design}
\label{sec:SystemDesign}
Theoretically, all the aforementioned architectures are feasible. 
In making a decision, it is important to consider the pertinent system requirements along with the foundational use case to guide the choice.

\subsection{Quality Criteria}
\label{subsec:SystemDesign-Quality}
The choice of architecture has a significant impact on the characteristics and quality criteria of the system. 
Functional quality criteria (cf. ISO/IEC~25000~\cite{ISO25000}) are reliability, usability, performance, security, supportability and transferability.
Quality criteria are evaluated from the perspective of a software provider that provides AAS instances to customers. 
For example, a component manufacturer that provides the descriptive model and, depending on the architecture, the service component within an AAS for a suitable physical asset. 
In general, the more software the AAS contains, the more elaborate it is for the software provider. 
\autoref{tab:Evaluation} shows the evaluation of the architectures mentioned above.

Reliability considers the tolerance to failure and recovery of a system. 
Since the evaluation focuses on the AAS, Lvl.~0 and Lvl.~1 can be used without hesitation. At these levels, the AAS consists only of the descriptive model, which is fault-tolerant due to the lack of logic.
Lvl.~2 has initial logic for processing the parameterization of requests. 
Problems within the AAS could affect requests and thus also data exchange. 
This means that reliability must at least be considered in this case.
The AASs in Lvl.~3 to Lvl.~5 have complex software components and are therefore responsible for a large part of the processing logic. 
Due to the increasing complexity and dependencies on essential software components within the AAS, reliability is becoming increasingly important. 
Lvl.~5 is the least critical due to the execution of a single application file, while at Lvl.~3 and Lvl.~4, the source code must still be processed before being executed. 
For these reasons, reliability must be given special consideration at these architecture levels.

Usability refers to the comprehensibility, learnability, operability, and attractiveness of a system. 
The question here is how easy it can be to build and operate the AAS. 
Lvl.~0 can be used without hesitation in this context, as a static model is easy to understand and operate. 
Lvl.~1 and Lvl.~2 are more difficult to implement due to the existing interface for data exchange and the simple logic within the AAS. 
In order to process the parameters in Lvl.~2, the corresponding knowledge must be applied within the AAS.
Lvl.~3 to Lvl.~5 require careful consideration, as complex logic in the form of source code or applications is used within the AAS.
This makes it more difficult to familiarize oneself with the system and operate it.
In general, it can be assumed that the greater the complexity of the software system within the AAS, the more challenging its development gets, and the more emphasis must be placed on ensuring usability.

The same categorization as for usability was used for performance. 
At Lvl.~0, performance does not need to be considered, as no runtime is expected with a static model. 
In Lvl.~1 and Lvl.~2, data requests must be processed internally, which allows basic logic and thus also initial optimizations to be carried out during execution. 
Low call times can have a positive effect on system performance and reduce block times. 
Due to the high software complexity within the AAS in Lvl.~3 to Lvl.~5, there is a significant need for optimization here. 
Since a large part of the logic is now located within the AAS, the runtime must be kept correspondingly low. 
Performance must be taken into account during AAS planning.

It is important to know where the system has the corresponding vulnerabilities. 
The various levels are evaluated using the security criterion.
Lvl.~0 does not need to be considered in this context, as a static model does not introduce any executable logic and therefore does not introduce vulnerabilities into the system, except for the possibility of the model itself being stolen.
Lvl.~1 and Lvl.~2 are partially compromised by requests via the API and the resulting data exchange.
In Lvl.~1, network attacks such as man-in-the-middle attacks could result in entire data sets being stolen. Man-in-the-middle attacks link into the data communication between two parties and can secretly sniff, change, and even delete data. This means that secure data exchange is no longer possible~\cite{conti2016}.
Through parameterization at Lvl.~2, SQL injections can be used in addition to network attacks to compromise queries in data requests, thus allowing specific data to be retrieved illegally. 
Lvl.~3 to Lvl.~5 are vulnerable to security breaches due to their complex software components. 
In addition to malicious code, dependencies and external libraries pose potential risks.
Simpler structures with few dependencies (cf. Lvl.~3) are preferable to more complex structures with stronger dependencies (cf. Lvl.~4). 
Integration of a black box (cf. Lvl.~5) should be prevented whenever possible, as this prevents analyzing the underlying code. 
Malware can lead to various worst-case scenarios, such as data theft, data loss, or complete system takeover.
Therefore, security should always be considered during implementation.

Supportability describes the analyzability, modifiability, stability, and testability of systems.
From the AAS perspective, the more internal software available within the AAS, the more difficult it is to modify.
The static model within Lvl.~0 can be easily adapted due to the lack of logic and external dependencies. 
The same applies to Lvl.~1, as no internal logic must be adapted here either, despite the API connection. 
In Lvl.~2, only parameterization exists as an internal AAS logic and adaptations are easy to perform.
Since the functionality of Lvl.~3 and Lvl.~4 is integrated into the AAS, this makes changes and tests more difficult. 
Changes would have to be adapted individually in each AAS. 
Since the source code for the logic is available within the AAS, changes and more in-depth testing are possible, although they are time-consuming. 
These conditions must be taken into account, at least in part, during implementation. 
Lvl.~5 can only be analyzed superficially and is difficult to modify. 
The executable file within the AAS must be treated as a black box, which means that only superficial tests are possible and no modifications can be made. 
For adjustments, the source code must first be changed outside of the AAS, a new system file must be generated from this, and has to be updated in the AAS.

The final criterion is transferability, which describes the effort required to transfer a component to another system.
Lvl.~0 and Lvl.~5 can be used without hesitation with respect to this criterion. 
Lvl.~0 has no logic and, therefore, no external dependencies, and Lvl.~5 already has the required logic as an executable file within the AAS. 
Both are therefore system-independent.
Like Lvl.~5, Lvl.~3 and Lvl.~4 also have the necessary logic internally in the AAS, but depend on an interpreter or compiler from the runtime to be executed. 
The severity of this depends on how the AAS was implemented. 
If the AAS only contains the static model and the source code data, then the system is responsible for loading all libraries and satisfying dependencies on external code. 
If, in addition to the data mentioned above, the AAS also has a virtual environment or a container with the necessary dependencies, this decouples the dependency on the system to a certain extent.
Lvl.~1 and Lvl.~2 have external dependencies due to data retrieval through the API. 
The requests require functioning routing with the corresponding IP addresses and host names. These are system-specific, but must be known to the AAS in order to receive the messages. This means that a simple transfer to another system is not possible, as these parameters must be adjusted. 
These dependencies must be taken into account when implementing the system.

In summary, the above criteria must be considered when making architectural decisions and then taken into account in the implementation of the system. 
Depending on the point of view for the evaluation of the criteria, different results need to be expected.

\begin{table}
    \centering
    \caption{Assessment of system architectures applying ISO/IEC~25000}
    \footnotesize
    \label{tab:Evaluation}
    \begin{tabular}{|c|c|c|c|c|c|c|}
        \hline
%        \multirow{2}{*}{ } &
%           \multicolumn{5}{c|}{Stages of software-heavy AASs} \\
           & Lvl.~0 & Lvl.~1 & Lvl.~2 & Lvl.~3 & Lvl.~4 & Lvl.~5\\
        \hline
        Reliability  & \colorcircle{green} & \colorcircle{green} & \colorcircle{orange} & \colorcircle{red} & \colorcircle{red} & \colorcircle{red}\\
        Usability & \colorcircle{green} & \colorcircle{orange} & \colorcircle{orange} & \colorcircle{red} & \colorcircle{red} & \colorcircle{red}\\
        Performance  & \colorcircle{green} & \colorcircle{orange} & \colorcircle{orange} & \colorcircle{red} & \colorcircle{red} & \colorcircle{red}\\
        Security & \colorcircle{green} & \colorcircle{orange} & \colorcircle{orange} & \colorcircle{red} & \colorcircle{red} & \colorcircle{red}\\
        Supportability & \colorcircle{green} & \colorcircle{green} & \colorcircle{green} & \colorcircle{orange} & \colorcircle{orange} & \colorcircle{red}\\
        Transferability & \colorcircle{green} & \colorcircle{red} & \colorcircle{red} & \colorcircle{orange} & \colorcircle{orange} & \colorcircle{green}\\
        \hline
        \multicolumn {6}{l}{
        \begin{tabular}{l}
            \footnotesize \colorcircle{green} without hesitation  \\ 
            \footnotesize \colorcircle{orange}  partly to be considered  \\
            \footnotesize \colorcircle{red} strongly to be considered  \\
        \end{tabular}
        }
    \end{tabular}
\end{table}

\subsection{Use Cases}
\label{subsec:SystemDesign-UseCases}
The usefulness of each architecture strongly depends on the specific use case and business model.
A common goal is to retrieve data about a physical asset from the customer's system. 
Various providers need the data for purposes such as preparing quotations or for internal processing.
In order to provide an API for data retrieval, the appropriate infrastructure and data connection are required.
From the customer's point of view, the data connection can be complex to provide depending on the chosen architecture.
At Lvl.~1, this connection would result in a great deal of effort, as a parameterized API endpoint would be required for each AAS. 
This would also make it difficult for the system to scale to a large number of AAS.
Since parameterization has been moved from the API to the request in Lvl.~2, a single interface can be used for the AAS. 
Reusability would keep the effort low and make the system easier to maintain and scale.
It should be noted that outsourcing parameterization also results in a certain loss of control on the user's part. Depending on the application, it is necessary to weigh up how strictly requests within the system should be regulated in order to prevent or hinder misuse. 
Consequently, a cost-benefit analysis should be performed.

The integration of logic within the AAS, as described in  Lvl.~3 to 5, is useful when the provider is not interested in the underlying data and when there is no need to consolidate data from different sources. 
Some models may require data from multiple machines, making data differentiation essential. 
Although setting up each AAS individually requires effort, this can be compensated by using standard algorithms in Lvl.~3 and Lvl.~4, with the option to extend them through custom logic.

In Lvl.~3 and Lvl.~4, the source code is accessible~(\textit{white~box}), enabling transparency, debugging, and modification.
In contrast, Lvl.~5 contains only one executable file~(\textit{black~box}), hiding implementation details. 
Since the structures of Lvl.~3 to Lvl.~5 are similar, choosing the appropriate level depends on the context. 
Lvl.~4, while allowing runtime code injections and dynamic adaptation, poses security and traceability risks and should mainly be used for research or testing.
For customer-side deployment, Lvl.~5 is recommended due to easier integration and bundled dependencies. 
For internal provider use, Lvl.~3 or Lvl.~4 may be preferable because accessible source code simplifies maintenance and customization. 
The choice between scripting (cf. Lvl. 3) and full services (cf. Lvl. 4) should depend on task complexity—simple calculations may suit scripts, while complex tasks require full services. 
Agent systems and peer-to-peer AAS interaction (cf.~VDI/VDE~2193~\cite{VDI2193}) further increase the relevance of Lvl.~3 to Lvl.~5, as embedded services enable more efficient and less complex agent communication.
In summary, it is important to first identify and define the use case before selecting an AAS architecture. 
This is the only way to ensure that the advantages and disadvantages of the selected architecture are optimally suited to the use case.

%---------------------------------------------------------------------------------------------------
% 05_Conclusion
%---------------------------------------------------------------------------------------------------
\section{Conclusion}
\label{sec:Conclusion}

This paper introduced a classification for software heaviness in AAS based on the runtime environment and functionality, evaluating its impact on system architectures, quality criteria, and use cases.
The selected level of software heaviness should not be static, but depending on use-cases that may evolve with the possibility to increase and decrease functional complexity.
The functionality must be selected and developed, in the initial conceptualization, as well as reevaluations, in accordance with the runtime environment and resulting quality criteria.
Future research should focus on developing methods to manage this lifecycle adaptability, reinforcing the AAS's role as a flexible digital twin.

%\vfill\pagebreak

\section*{Acknowledgements}

Partly funded by the Federal Ministry for Economic Affairs and Energy (BMWE) through the project growING (grant no. 13IPC036G).
Partly funded by the German Federal Ministry of Research, Technology and Space (BMFTR) within the ``Research Campus – Public-Private Partnership for Innovation'' funding initiative (02P23Q820) and managed by the Project Management Agency Karlsruhe (PTKA).

\bibliographystyle{elsarticle-num}
\bibliography{bibliography}

%\clearpage\onecolumn

\end{document}